\documentclass[11pt]{article}

\usepackage{hyperref}
\usepackage{url}
\usepackage{graphicx}
\usepackage{caption}
\usepackage{subcaption}   
\usepackage{geometry}
\usepackage{makecell}
\usepackage{booktabs}
\usepackage{multirow}
\usepackage{amsmath}
\usepackage{array}
\usepackage{tabularx}
\usepackage[table]{xcolor}
\usepackage{listings}
\usepackage{float}
\usepackage{multicol}
\usepackage{enumitem}
\usepackage{tasks}
\usepackage{hyperref}

\usepackage[preprint]{acl}

\usepackage{times}
\usepackage{latexsym}

\usepackage[T1]{fontenc}

\usepackage[utf8]{inputenc}

\usepackage{microtype}

\usepackage{inconsolata}

\usepackage{graphicx}

\title{Multi-Docker-Eval: A `Shovel of the Gold Rush' Benchmark on Automatic Environment Building for Software Engineering}

\author{
    \textbf{Kelin Fu\textsuperscript{1,\dag}},
    \textbf{Tianyu Liu\textsuperscript{1,*}},
    \textbf{Zeyu Shang\textsuperscript{2}},
    \textbf{Yingwei Ma\textsuperscript{2}},
    \\
    \textbf{Jian Yang\textsuperscript{3}}, 
    \textbf{Jiaheng Liu\textsuperscript{3}}, 
    \textbf{Kaigui Bian\textsuperscript{1,\dag}}
    \\
    \textsuperscript{1}School of Computer Science, Peking University \\
    \textsuperscript{2}Moonshot AI \:
    \textsuperscript{3}M-A-P\\
  \small{
    \textsuperscript{\dag}\textbf{Correspondence:} 
    litble@stu.pku.edu.cn, \{tianyu0421, bkg\}@pku.edu.cn \* \textsuperscript{*}Project Leader
  }
  \\
  \raisebox{-1.5pt}{\includegraphics[height=1.05em]{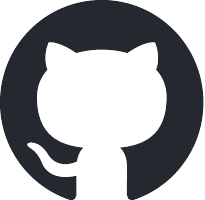}}~\texttt{GitHub:}
  \url{https://github.com/Z2sJ4t/Multi-Docker-Eval}
  \\
\raisebox{-2.5pt}{\includegraphics[height=1.05em]{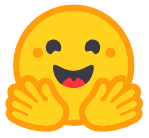}}~\texttt{HuggingFace:}
  \url{https://huggingface.co/datasets/litble/Multi-Docker-Eval}\\ \\
}

\begin{document}
\setlist[itemize]{noitemsep, topsep=0pt, parsep=0pt, partopsep=0pt}

\maketitle
\begin{abstract}

Automated environment configuration is a critical bottleneck in scaling software engineering (SWE) automation. 
To provide a reliable evaluation standard for this task, we present Multi-Docker-Eval benchmark. It includes 40 real-world repositories spanning 9 programming languages and measures both success in achieving executable states and efficiency under realistic constraints. Our extensive evaluation of state-of-the-art LLMs and agent frameworks reveals key insights: (1) the overall success rate of current models is low (F2P at most 37.7\%), with environment construction being the primary bottleneck; (2) model size and reasoning length are not decisive factors, and open-source models like DeepSeek-V3.1 and Kimi-K2 are competitive in both efficiency and effectiveness; (3) agent framework and programming language also have significantly influence on success rate. These findings provide actionable guidelines for building scalable, fully automated SWE pipelines.
\end{abstract}

\section{Introduction}

Software repositories on platforms like GitHub underpin modern software engineering, enabling large-scale collaboration and open-source development~\citep{swebench, fan2023large}. Among the numerous challenges in this space, repository-level tasks---such as bug-fixing, optimization and adding new features---are of particular importance. Solving these tasks requires a deep understanding of codebases, dependencies, and build environments. Recently, Large Language Models (LLMs) have shown promising capabilities in tackling such problems, advancing the frontier of automated issue resolution and repository-level code generation.

A typical repository-level problem can be formulated as: given a repository $R$ and a test function $T(\cdot)$, the goal is to generate a patch $P$ such that $T(R)$ fails initially and $T(R\oplus P)$ passes after applying $P$. Success hinges on constructing executable environments that accurately reflect repository states before and after patching.

However, building such environments remains challenging due to diverse dependencies, language versions, and build configurations. While prior efforts like SWE-Gym~\citep{swe_gym} and swe-rebench~\citep{swe_rebench} used manual or rule-based setups, and others like SWE-Smith~\citep{swe_smith} and R2E-Gym~\citep{r2e_gym} employed synthetic data, scaling environment configuration is still difficult. Recent agentic systems (e.g., SWE-Factory~\citep{swe_factory}, RepoLaunch~\citep{repolaunch}) address this by automating Docker-based environment setup, enabling scalable and continuous data generation for training and evaluation.

We analogize this process to a modern-day ``gold rush'': repositories are gold mines, automated fixes are the gold, and environment-configuring agents are the shovels. A reliable ``shovel'' is essential for efficient extraction. To evaluate these tools, we introduce Multi-Docker-Eval, a benchmark for assessing automated environment configuration across language ecosystems.

Existing benchmarks like EnvBench~\citep{envbench} and INSTALLAMATIC-bench
~\citep{beyond_pip} are limited: they often measure only setup success, without verifying test validity or task-specific correctness. Most are language-specific (e.g., Python or JVM) and use narrow metrics, ignoring configuration time, resource use, and test robustness—key factors for real-world use.

To address these gaps, we present Multi-Docker-Eval, a multi-language, multi-dimensional evaluation framework for automatic environment construction. It assesses not only the success of achieving executable states, but also the efficiency and stability of the configuration process under realistic resource constraints. We envision Multi-Docker-Eval as a benchmark ``shovel test''---a practical and principled step toward empowering the next generation of data-driven software intelligence.

Our contributions are as follows:

\begin{itemize}
    \item \textbf{Multi-Docker-Eval benchmark}: The first multilingual benchmark for evaluating LLM agents on configuring executable environments and test scripts for real-world repositories.
    \item \textbf{Comprehensive model evaluation}: Extensive experiments with open- and closed-source LLMs to assess capability and efficiency, guiding future automated data generation.
    \item \textbf{Framework comparison}: Analysis of SWE-Builder and RepoLaunch, summarizing strengths and weaknesses to inform framework design.
    \item \textbf{Toward full automation}: Design insights for scalable, cost-effective, and fully automated pipelines in software engineering benchmark and dataset construction.
\end{itemize}

\section{Related Works}

\textbf{Repo-level coding tasks}. Recent benchmarks assess LLMs in resolving real-world GitHub issues. SWE-bench~\citep{swebench}, the most widely used benchmark, provides Python issues paired with test suites. Extensions like R2E-Gym~\citep{r2e_gym} and SWE-Smith~\citep{swe_smith} expand the task scope using LLM-synthesized data, while swebench-multilingual~\citep{swe_smith} and Multi-SWE-bench~\citep{multi-swe-bench} extend evaluation to multilingual settings. Beyond bug-fixing, recent benchmarks assess broader capabilities: nocode-bench~\citep{nocode_bench} and FEA-Bench~\citep{fea_bench} targets feature addition, GSO-bench~\citep{gso_bench} focuses on code optimization, SWTbench~\citep{SWTbench} evaluates unit test generation, and long-horizon tasks are explored in Commit0~\citep{commit0} and SWE-bench Pro~\citep{swebench_pro}. These efforts collectively advance automated issue resolution and repository-level code generation using LLMs~\citep{agentless, swe_agent, alibaba-lingma-agent, swegpt}.

\textbf{Agent frameworks for automated repository setup}. To automate the complex setup process, several agent frameworks have emerged. ExecutionAgent~\citep{bouzenia2025you} generates scripts for building and testing projects. RepoLaunch~\citep{repolaunch} and SetUpAgent~\citep{setup_agent} employ bash-interactive agents
. Repo2Run~\citep{repo2run} hits 86\% configuration success rate in Python. Advancing further, SWE-Factory~\citep{swe_factory} introduces a multi-agent collaboration framework that achieves fully automated, multi-language environment configuration.

\textbf{Benchmarks for environment configuration}. Efforts to evaluate automated configuration include EXECUTIONAGENT-bench~\citep{bouzenia2025you}, which assesses build and test rates across 10 repositories in 5 languages but requires manual scoring. INSTALLAMATIC-bench
~\citep{beyond_pip} focuses on Python, and EnvBench~\citep{envbench} extends to JVM languages. However, these benchmarks do not evaluate configuration for specific repository issues nor effectively test the agent's ability to generate corresponding test scripts.

\section{Multi-Docker-Eval Benchmark}

In this chapter, we descirbe Multi-Docker-Eval, our benchmark designed for automated executable environment configuration.

\subsection{Task Definition}

\begin{figure}
    \centering
    \includegraphics[width=0.99\linewidth]{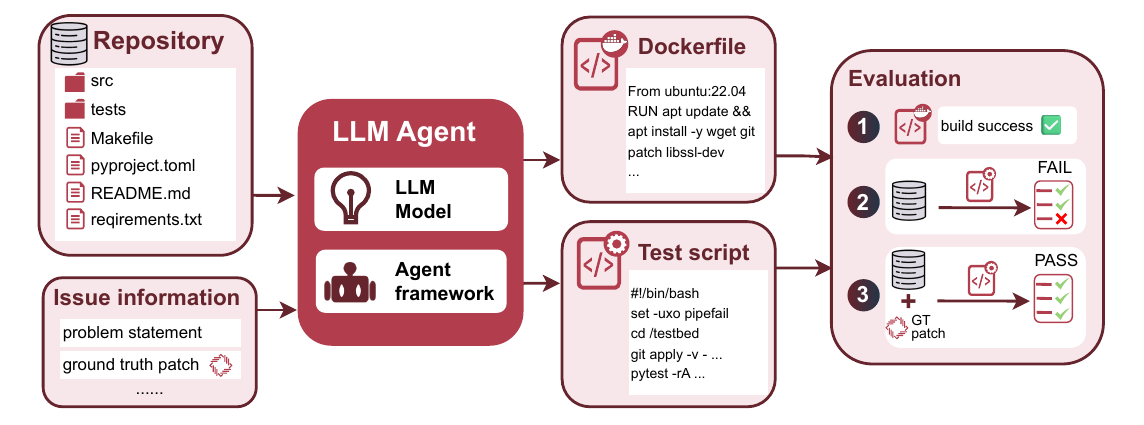}
    \caption{Overview of the workflow with Multi-Docker-Eval.}
    \label{fig:main}
    \vspace{-0.15in}
\end{figure}

As shown in Figure~\ref{fig:main}, the Multi-Docker-Eval benchmark assesses a model's capacity to automate environment creation and testing for repository-level code tasks. The process begins with an input triplet $<R, D, P^*>$: a source repository $R$, a natural language problem description $D$, and the correct solution patch $P^*$. The model must then produce two outputs: a test function $T(\cdot)$ that fails on the original code $R$ but passes on the patched version $R \oplus P^*$, and a runnable Docker environment to execute this test. The benchmark ultimately evaluates the agent's ability to install dependencies and generate valid tests from problem descriptions.

\subsection{Data Collection and Filtering}
\label{sec:data_collect}

\begin{figure*}[htbp]
    \centering
    \begin{subfigure}[b]{0.24\linewidth}
      \centering
      \includegraphics[width=\linewidth]{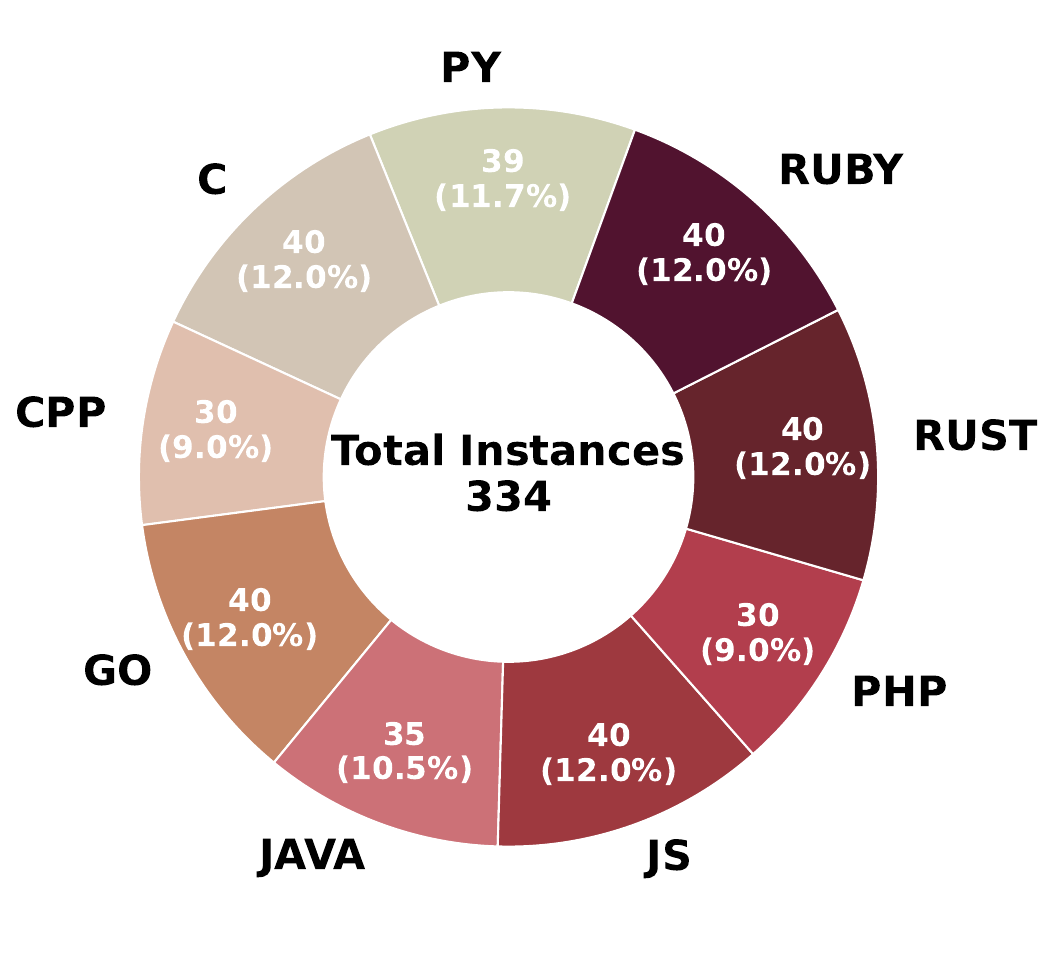}
      \caption{Instances by language}
      \label{fig:distribution1}
    \end{subfigure}\hfill
    \begin{subfigure}[b]{0.22\linewidth}
      \centering
      \includegraphics[width=\linewidth]{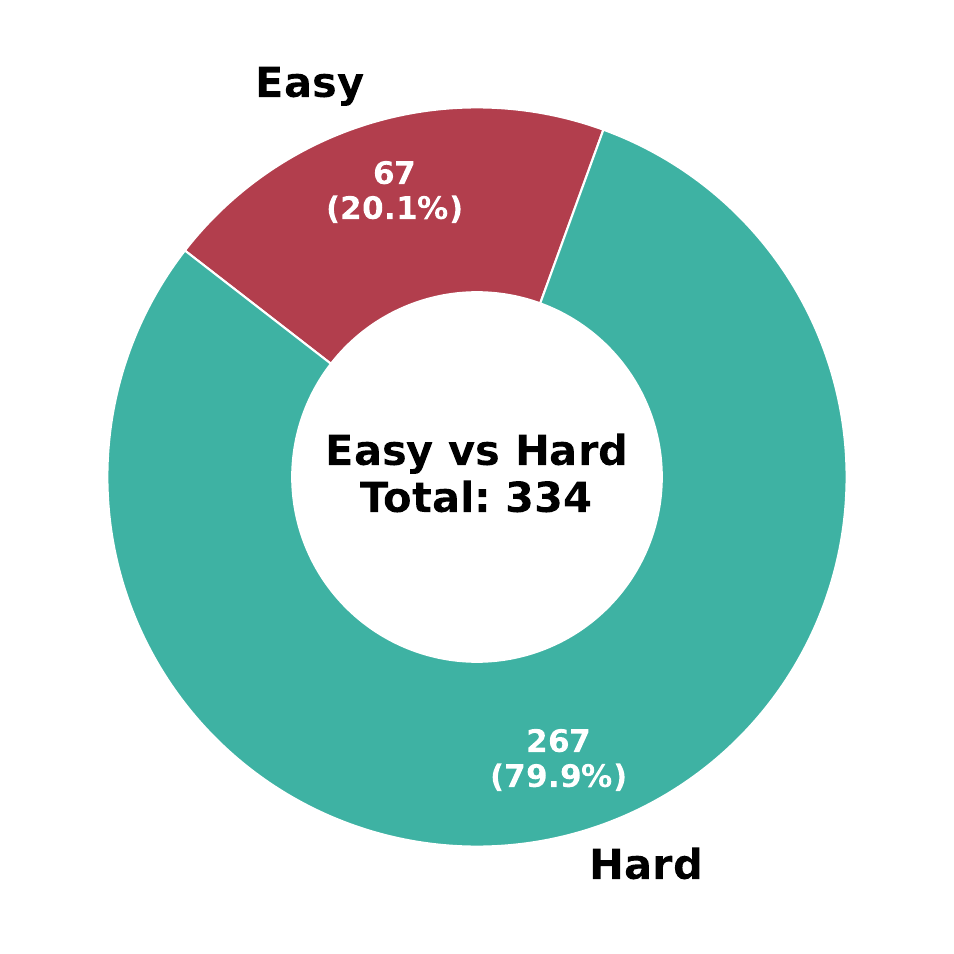}
      \caption{Difficulty split}
      \label{fig:distribution2}
    \end{subfigure}\hfill
    \begin{subfigure}[b]{0.23\linewidth}
      \centering
      \includegraphics[width=\linewidth]{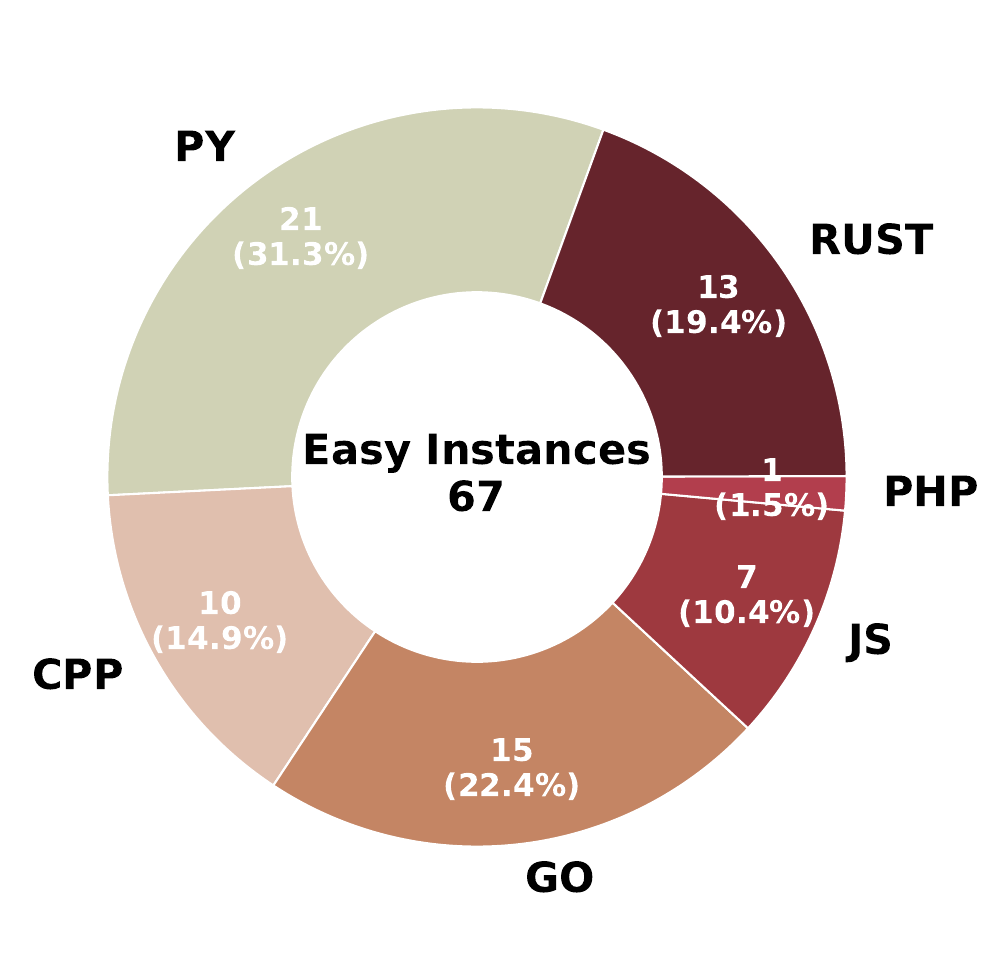}
      \caption{Easy cases by language}
      \label{fig:distribution3}
    \end{subfigure}\hfill
    \begin{subfigure}[b]{0.24\linewidth}
      \centering
      \includegraphics[width=\linewidth]{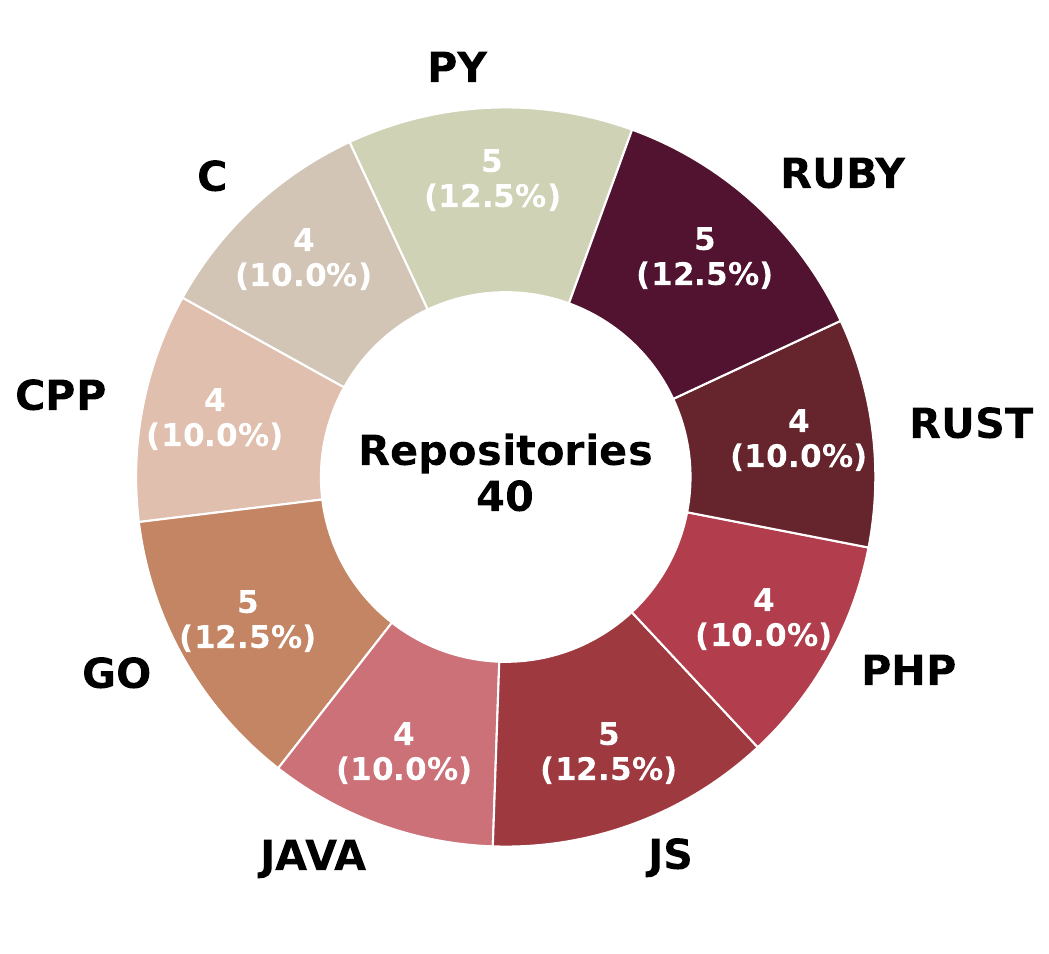}
      \caption{Repositories by language}
      \label{fig:distribution4}
    \end{subfigure}
    
    \caption{Overview of the data composition of Multi-Docker-Eval.}
    \label{fig:distribution}
    \vspace{-0.15in}
\end{figure*}

\textbf{Dataset construction}. As detailed in Figure~\ref{fig:distribution1} and Figure~\ref{fig:distribution4}, We collected 334 issues from 40 open-source repositories across 9 popular programming languages from GitHub (for specific repository information, please refer to Appendix~\ref{app:repo}). These languages cover major programming paradigms and are among the top 20 most searched-for on Google~\citep{pypl_index}.

\textbf{Filtering}. To ensure quality and avoid overly popular repositories, we selected those meeting:  (i) 1000 $\leq$ stars $\leq$ 1500 (ii) forks $\geq$ 20 (iii) contributors $\geq$ 10 (iv) repository size $\leq$ 100 MB. For each repository, we select at most 8 pull requests created after 31 July 2025 (UTC).

\textbf{Difficulty verification}. For each repository version, we verified if a runnable environment could be easily configured (e.g., via \texttt{pip install -r requirements.txt}). Configurations achievable this way are labeled ``Easy''; others are ``Hard''. In our dataset, 20.06\% (67 instances) are ``Easy''. The specific difficulty distribution of Multi-Docker-Eval can be found in Figure~\ref{fig:distribution2} and Figure~\ref{fig:distribution3}. The specific commands are provided in Appendix~\ref{app:diffculty_verf}.

\subsection{Metrics}
We introduce two metric categories in Multi-Docker-Eval: outcome and process metrics.

\textbf{Outcome metrics} assess the correctness of the environment and tests:
\begin{itemize}
\item \textbf{Fail-to-pass rate (F2P)}: The proportion of tests that fail before and pass after configuration. This is the main metric.
\item \textbf{Commit rate}: The proportion of instances where the agent commits its output for evaluation.
\end{itemize}

\textbf{Process metrics} evaluate efficiency and resource usage:
\begin{itemize}
\item \textbf{Token consumption}: Total tokens used by LLMs during configuration.
\item \textbf{Wall time}: Total real time from configuration start to completion.
\item \textbf{CPU seconds}: Total CPU time used.
\item \textbf{Max Resident Set Size (Max RSS)}: Peak memory usage.
\item \textbf{Average image size}: Average size of final Docker images, indicating storage efficiency.
\end{itemize}

\section{Experimental Setup}
\label{sec:experimental_setup}

\textbf{Agent framework}. We evaluated two agent frameworks for automated environment configuration. Detailed introductions are provided in Appendix~\ref{app:framework}:
\begin{itemize}
    \item \textbf{SWE-Builder (multi-agent)}: four specialised agents jointly search the repo, write the Dockerfile, install dependencies and generate tests; a built-in loop re-invokes the agents when a test fails, and a memory pool reuses previously validated environments.
    \item \textbf{RepoLaunch (single agent)}: one agent sequentially picks a base image and issues raw bash commands inside the container; no automatic retry or reuse.
\end{itemize}

The former trades flexibility for stability; the latter offers an open-ended command space but suffers from higher variance.

\textbf{LLM models}. 7 open-source (DeepSeek-v3.1, DeepSeek-R1, Qwen3-235B-A22B, GPT-OSS-20/120B, Kimi-K2-0905, Kimi-K2-thinking) and 3 closed-source (Claude-Sonnet-4, GPT-5-Mini, Gemini-2.5-Flash).

\section{Results}

\begin{table*}[thbp]
  \centering
  \footnotesize
  \setlength{\tabcolsep}{5.5pt}
  \begin{tabular}{@{}lcccccccc@{}}
    \toprule
    \multicolumn{1}{c}{\multirow{2}{*}{\textbf{Model}}} & \multicolumn{2}{c}{\textbf{Outcome Metrics}} & \multicolumn{6}{c}{\textbf{Process Metrics}} \\
    \cmidrule(lr){2-3} \cmidrule(lr){4-9}
     & \makecell{F2P\\(\%)} & \makecell{Commit \\ (\%)} & \makecell{Avg \\input\\tokens (K)} & \makecell{Avg \\output\\tokens (K)} & \makecell{Wall \\time\\(Ks)} & \makecell{CPU \\seconds \\ (Ks)} & \makecell{Max\\ RSS\\(GB)} & \makecell{Avg \\docker\\size (GB)} \\
    \addlinespace
    \hline
    \rowcolor{gray!30}\multicolumn{9}{c}{\textit{Open-source Models}} \\
    \texttt{DeepSeek-v3.1} & \textbf{\underline{37.72}} & 52.89 & 158.11 & 17.15 & 122.80 & 0.749 & 7.45 & 1.02 \\
    \rowcolor{gray!10}\texttt{DeepSeek-R1} & 26.65 & 41.72 & 138.05 & 60.10 & 143.37 & 0.534 & 7.47 & 1.02 \\
    \texttt{GPT-OSS-20B} & 17.17 & 29.44 & 184.23 & 58.37 & 127.91 & 0.626 & 7.38 & 0.87 \\
    \rowcolor{gray!10}\texttt{GPT-OSS-120B} & 27.00 & 37.72 & 128.31 & 30.17 & 104.29 & 0.478 & 8.70 & 0.83 \\
    \texttt{Kimi-K2-0905} & 37.62 & 55.49 & 113.02 & 7.92 & 101.81 & 0.691 & 7.60 & 1.01 \\
    \rowcolor{gray!10}\texttt{Kimi-K2-thinking} & 36.53 & 52.69 & 162.12 & 59.40 & 114.61 & 0.425 & 7.66 & 1.05 \\
    \texttt{Qwen3-235B-A22B} & 23.65 & 34.53 & 101.46 & 38.87 & 138.64 & 0.613 & 7.38 & 1.00 \\
    \addlinespace
    \hline
    \rowcolor{gray!30}\multicolumn{9}{c}{\textit{Closed-source Models}} \\
    \texttt{Claude-Sonnet-4} & \textbf{\underline{35.53}} & 47.41 & 182.85 & 15.01 & 110.54 & 0.680 & 7.30 & 1.17 \\
    \rowcolor{gray!10}\texttt{GPT-5-Mini} & 34.13 & 49.60 & 339.94 & 103.32 & 158.61 & 0.746 & 7.34 & 0.95 \\
    \texttt{Gemini-2.5-Flash} & 29.44 & 40.62 & 153.43 & 32.60 & 88.00 & 0.698 & 7.58 & 0.97 \\
    \bottomrule
  \end{tabular}
  \caption{Overall Multi-Docker-Eval performance of different models on SWE-Builder.}
  \label{tab:main_result}
  \vspace{-0.15in}
\end{table*}

\subsection{RQ1: Effiency of state-of-the-art LLMs on Configuration Tasks}

To begin, we assess how effectively current LLMs perform the automated environment configuration task---that is, \textit{how well they convert a repository into an executable state with a valid test script (RQ1)}.

\textbf{Overall Performance}. As shown in Table~\ref{tab:main_result}, the primary metric, Fail-to-Pass Rate (F2P), ranges from 17\% to 38\% across models, indicating that fewer than half of repositories are successfully configured. This low success confirms that Multi-Docker-Eval remains challenging, requiring both dependency reasoning and complex engineering operations. The wide performance gap also suggests this capability is not yet broadly covered by pretraining or instruction-tuning data.

\begin{figure}[thbp]
    \centering
    \begin{subfigure}[b]{0.49\linewidth}
        \centering
        \includegraphics[width=\linewidth]{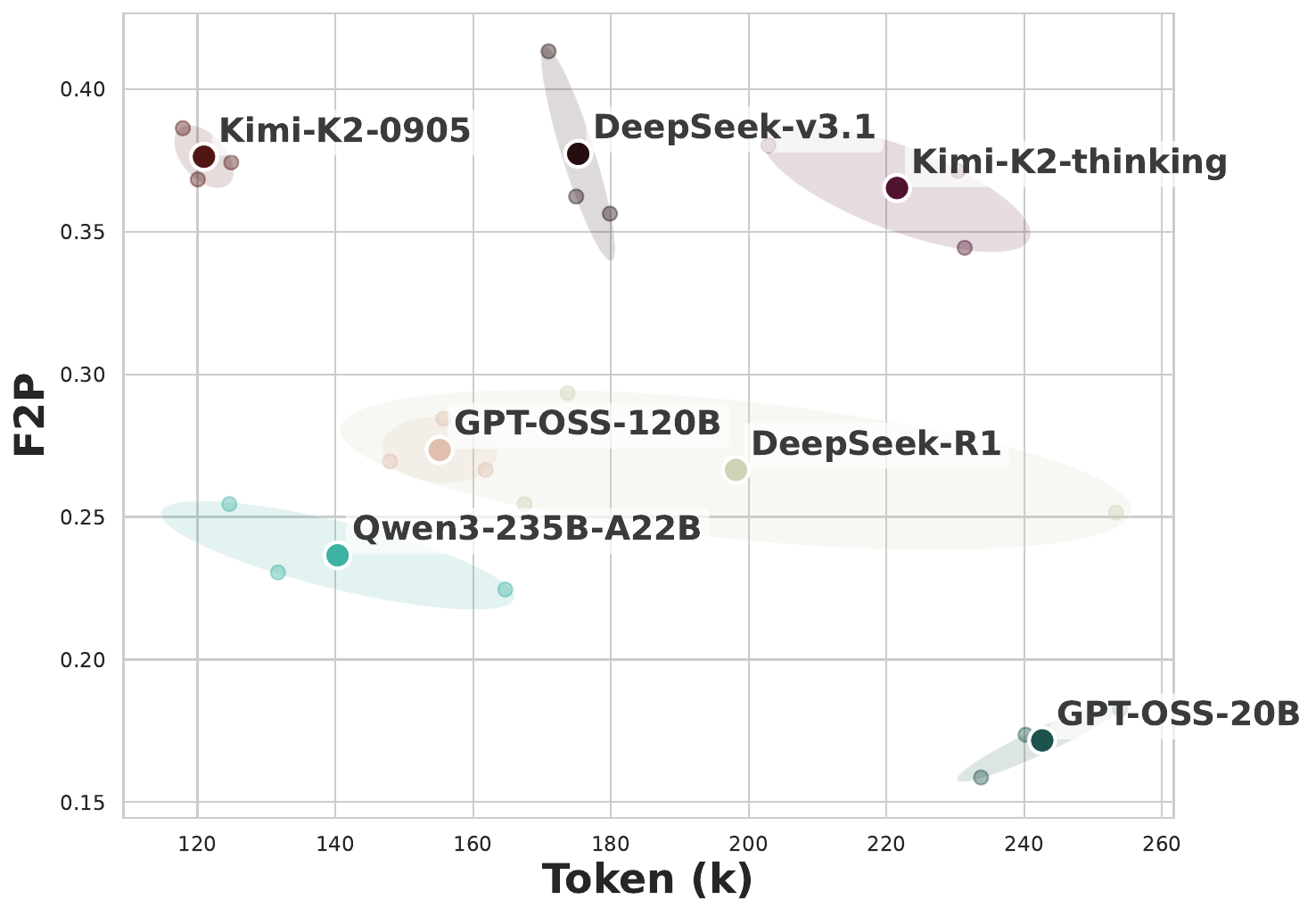}
        \caption{F2P vs Token}
        \label{fig:f2p_token}
    \end{subfigure}\hfill
    \begin{subfigure}[b]{0.48\linewidth}
        \centering
        \includegraphics[width=\linewidth]{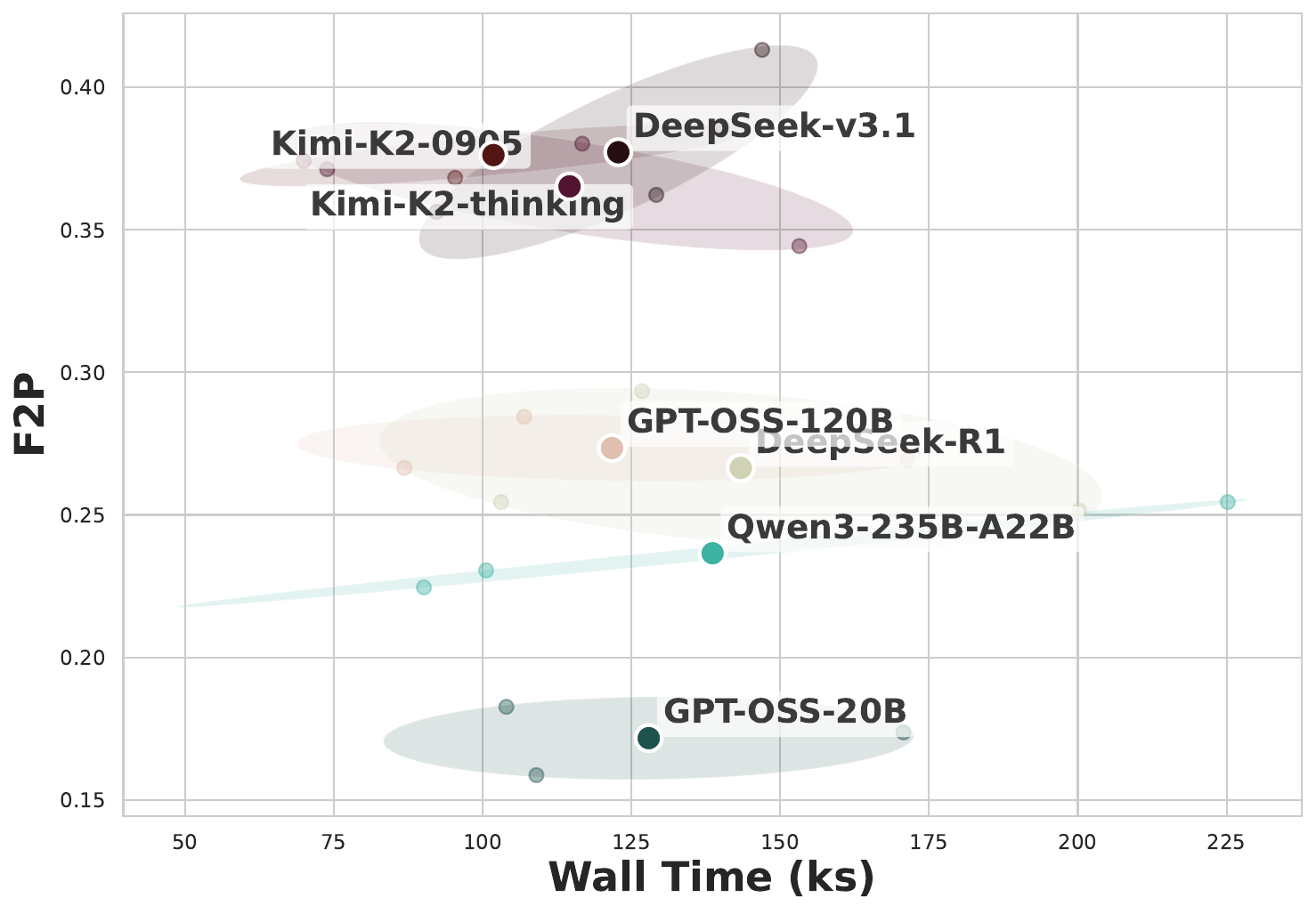}
        \caption{F2P vs Wall Time}
        \label{fig:f2p_walltime}
    \end{subfigure}
    \caption{Relationship between F2P and resource consumption metrics on different models. The scatter plot shows the results and average of our three repeated experiments.}
    \label{fig:resource_metrics}
    \vspace{-0.2in}
\end{figure}

\textbf{Model Comparison}. The open-source model \texttt{DeepSeek-v3.1} (37.72\%) achieves the highest F2P, outperforming \texttt{Claude-Sonnet-4} (35.53\%) and \texttt{GPT-5-Mini} (34.13\%). In terms of efficiency, as revealed in Figures~\ref{fig:resource_metrics}, \texttt{Kimi-K2-0905} attains 37.62\% F2P using notably fewer tokens ($\sim$ 120K) and less wall time (114.61s), whereas higher-performing models often demand more resources. For example, \texttt{GPT-5-Mini} consumes $\sim$443.26K tokens for an F2P of 34.13\%.

\textbf{Self-Evaluation Reliability}. The Commit Rate, reflecting an agent's confidence, often misaligns with actual performance. \texttt{Claude-Sonnet-4} commits 47.41\% of runs but achieves only 35.53\% F2P, and \texttt{Qwen3-235B-A22B} shows similar overconfidence. This indicates that LLM agents currently lack reliable self-assessment in complex software tasks.

\textbf{Resource and Efficiency Patterns}. All models require substantial wall time (roughly 24–44 hours per run), underscoring the task's engineering complexity. In contrast, CPU seconds, Max RSS, and Docker size remain stable—typically 600–750 CPU s, 7.4–7.7 GB RAM, and 0.9–1.2 GB image size—indicating that the SWE-Builder framework provides predictable, resource-bounded execution. This stability simplifies capacity planning for large-scale deployment.

In summary, \textbf{environment configuration performance does not directly correlate with model size or reasoning length}. Open-source models like \texttt{DeepSeek-v3.1} and \texttt{Kimi-K2-0905} match or exceed larger closed-source models with lower token costs. Moreover, SWE-Builder ensures consistent resource usage across models, allowing capacity planning to focus primarily on token budget and wall time.

\subsection{RQ2: Impact of Agent Framework Architectures}

\begin{table*}[htbp]
  \centering
  \small
  \begin{tabular}{@{}lcccccccc@{}}
    \toprule
    \textbf{Framework} & \makecell{F2P\\(\%)} & \makecell{Commit \\ (\%)} & \makecell{Avg \\input\\tokens (K)} & \makecell{Avg \\output\\tokens (K)} & \makecell{Wall \\time\\(Ks)} & \makecell{CPU \\seconds \\ (Ks)} & \makecell{Max\\ RSS\\(GB)} & \makecell{Avg \\docker\\size (GB)} \\
    \midrule
    SWE-Builder & \makecell{30.58 \\ ($\pm$ 6.54)} & \makecell{45.92 \\ ($\pm$ 8.18)} & \makecell{165.95 \\ ($\pm$ 63.59)} & \makecell{42.15 \\ ($\pm$ 27.27)} & \makecell{122.80 \\ ($\pm$ 19.63)} & \makecell{0.62 \\ ($\pm$ 0.11)} & \makecell{7.58 \\ ($\pm$ 0.09)}& \makecell{1.01 \\ ($\pm$ 0.09)}\\ \addlinespace\hline\addlinespace
    RepoLaunch   & \makecell{8.85 \\ ($\pm$ 3.11)} & \makecell{22.35 \\ ($\pm$ 9.04)} & \makecell{504.05 \\ ($\pm$ 465.94)} & \makecell{10.03 \\ ($\pm$ 8.07)} & \makecell{31.72 \\ ($\pm$ 10.34)} & \makecell{6.25 \\ ($\pm$ 13.78)} & \makecell{5.37 \\ ($\pm$ 1.60)}& \makecell{1.42 \\ ($\pm$ 0.09)}\\
    \bottomrule
  \end{tabular}
  \caption{Comparison of average performance metrics ($\pm$ standard deviation) between SWE-Builder and RepoLaunch frameworks among all 8 models.}
  \label{tab:framework_comparison}
  \vspace{-0.2in}
\end{table*}

Improving configuration quality also requires choosing the right agent architecture. Therefore, we next analyze \textit{how do different agent framework architectures impact the performance of environment configuration (RQ2)}. We compare two representative
frameworks---SWE-Builder and RepoLaunch, which has been introduced in Section~\ref{sec:experimental_setup}. Results across 10 models in Table~\ref{tab:framework_comparison} highlight clear distinctions in success rates, interaction efficiency, and resource usage patterns. More details of RepoLaunch experiment are provided in Appendix~\ref{app:repolaunch}.

\begin{figure}[htbp]
    \centering
    \begin{minipage}[b]{0.52\linewidth}
    \includegraphics[width=\linewidth]{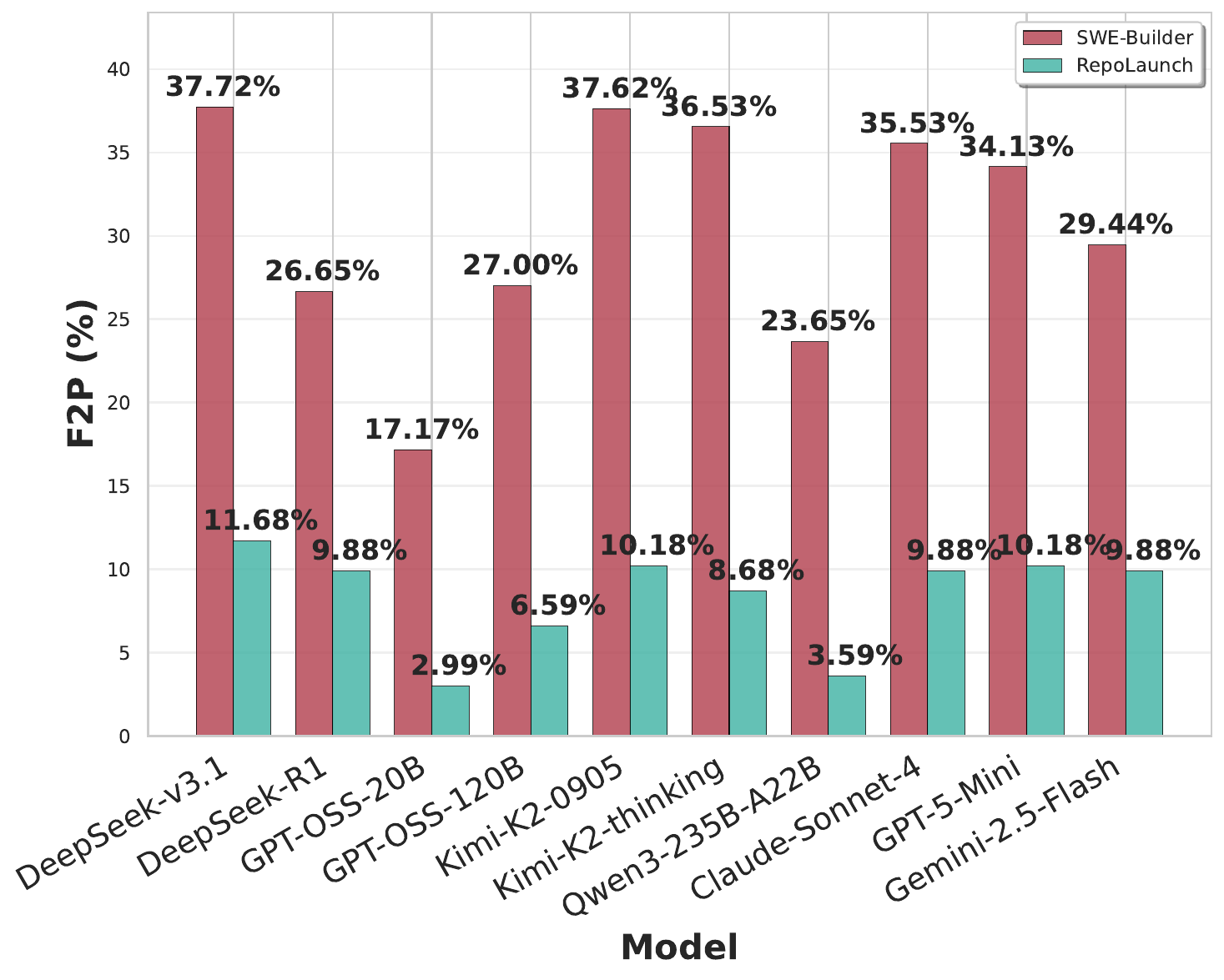}
    \caption{Variations in F2P across different models on SWE-Builder and RepoLaunch framework. Most models achieve F2P scores on SWE-Builder’s Multi-Docker-Eval that are 2.5- to 3.5-fold higher than those on RepoLaunch.}
    \label{fig:f2p_comp}
    \end{minipage}
    \hfill
    \begin{minipage}[b]{0.45\linewidth}
        \includegraphics[width=\linewidth]{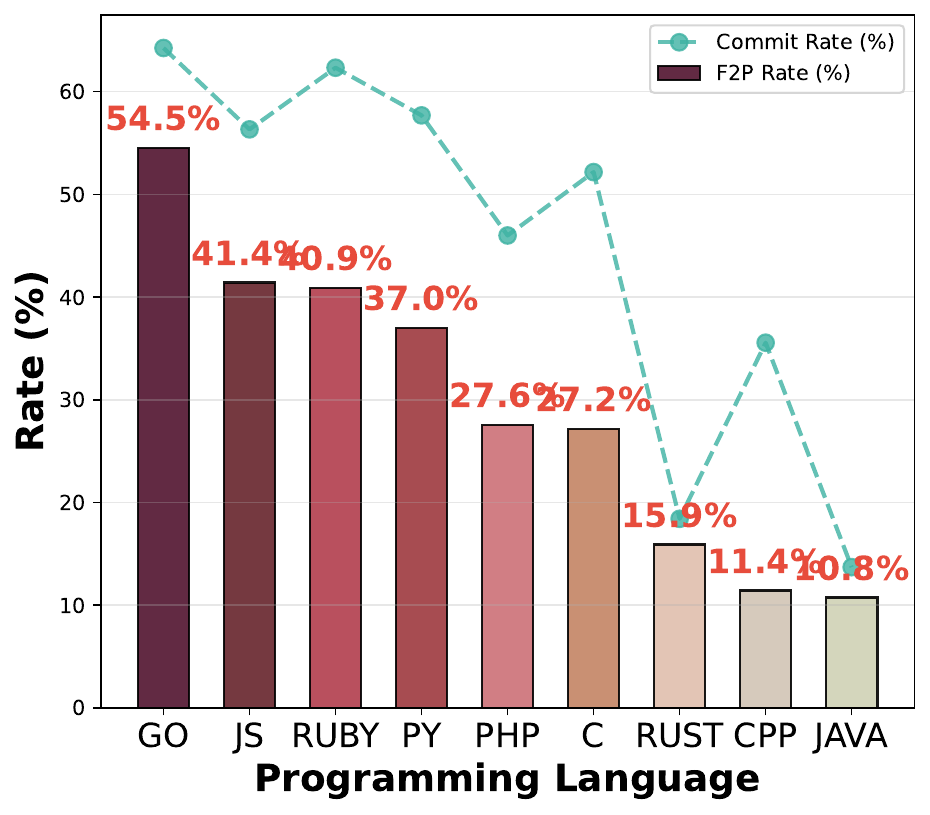}
    \caption{The bar chart shows the F2P rate across different programming languages, while the line graph indicates the commit rate (Framework: SWE-Builder).}
    \label{fig:language_hist}
    \end{minipage}
    \vspace{-0.15in}
\end{figure}

\textbf{Success Rate}. As shown in Figure~\ref{fig:f2p_comp}, RepoLaunch yields significantly lower F2P than SWE-Builder across all models. This gap stems from architectural differences: SWE-Builder uses four specialized agents that collaborate on exploration, setup, and testing, with reactivation capability for iterative error repair. In contrast, RepoLaunch relies on single-agent sequential reasoning, offering limited feedback or correction. Moreover, RepoLaunch’s commit rate (22.35\%) is notably higher than its F2P (8.85\%). This is due to its weaker internal judging mechanism: the framework often marks instances as complete even when the generated setup or test commands require manual inspection and rewriting to become usable. This highlights the lack of reliable automated validation in its original workflow.

\textbf{Divergent Runtime and Resource Patterns}. Process metrics reveal further contrasts:
\begin{itemize}
\item \textbf{Token Usage}: RepoLaunch uses 3.5$\times$ more input tokens (504.05K vs. 165.95K) due to accumulated bash history, but generates 4.2$\times$ fewer output tokens (10.03K vs. 42.15K), reflecting reactive command execution rather than structured planning.
\item \textbf{Dependency Efficiency}: RepoLaunch produces 41\% larger Docker images (1.41 GB vs. 1.01 GB), suggesting less optimized dependency resolution.
\item \textbf{Computational Profile}: RepoLaunch runs 3.9$\times$ faster in wall time (31.72Ks vs. 122.80Ks) but uses 10.08$\times$ more CPU time (6.25Ks vs. 0.62Ks) with high variance ($\pm$13.78Ks), indicating intensive yet unstable computation.
\end{itemize}

\textbf{Contrast in Resource Variance}. RepoLaunch exhibits high variance across token and CPU metrics, reflecting instability when reasoning with low-level bash commands. SWE-Builder, by contrast, uses declarative operations (e.g., repository inspection, configuration generation), leading to more predictable and reproducible resource usage.

In summary, framework architecture critically shapes automation reliability. Multi-agent collaboration with repair loops, as in SWE-Builder, supports higher success rates, while single-agent sequential designs amplify errors without self-correction. \textbf{A more effective ``shovel'' should thus adopt feedback-driven, memory-augmented workflows that enable iterative refinement and reuse of validated configurations.}

\subsection{RQ3: Variability of Configuration Success across Programming Language}

The programming language ecosystem is a major factor determining the difficulty of automated environment configuration. We thus examine \textit{how programming languages differ in configuration difficulty and model success (RQ3)}. As summarized in Figure~\ref{fig:language_hist} and Appendix~\ref{app:language}, several general trends can be observed as follows:

Languages with standardized, declarative build systems—such as Go, Python, and JavaScript—achieve the strongest results. Go performs best (Commit 64.25\%, F2P 54.50\%), reflecting its reproducible module system and minimal system-level dependencies. Python and JavaScript also excel, benefiting from mature package managers (\texttt{pip}, \texttt{npm}) and established conventions that guide LLMs toward correct setups. In contrast, C/C++, Java, and Rust exhibit higher failure rates, as their builds often depend on system libraries, compiler toolchains, and version-specific configurations that are difficult to infer.

Testing ecosystems further influence performance. Languages like Python and Go, with unified testing workflows (e.g., pytest, \texttt{go test}), enable reliable test script generation. However, PHP and JavaScript use diverse frameworks (e.g., PHPUnit, Jest, Mocha), complicating test entry point identification. PHP exemplifies this gap: its commit rate reaches 46.00\%, but its F2P drops to 27.56\%, indicating that models often misinterpret dependency installation as environment readiness, even when tests remain non-executable.

These findings highlight clear directions for improving configuration systems. Languages with uniform dependency management and testing workflows align well with current LLM capabilities. In contrast, ecosystems like C/C++, Java, Rust, and PHP reveal structural weaknesses in handling system dependencies, compiler setup, and fragmented test conventions. \textbf{These should be prioritized in future agent optimization.}

\subsection{RQ4: Bottleneck Analysis}

\begin{figure}
    \centering
    \includegraphics[width=0.9\linewidth]{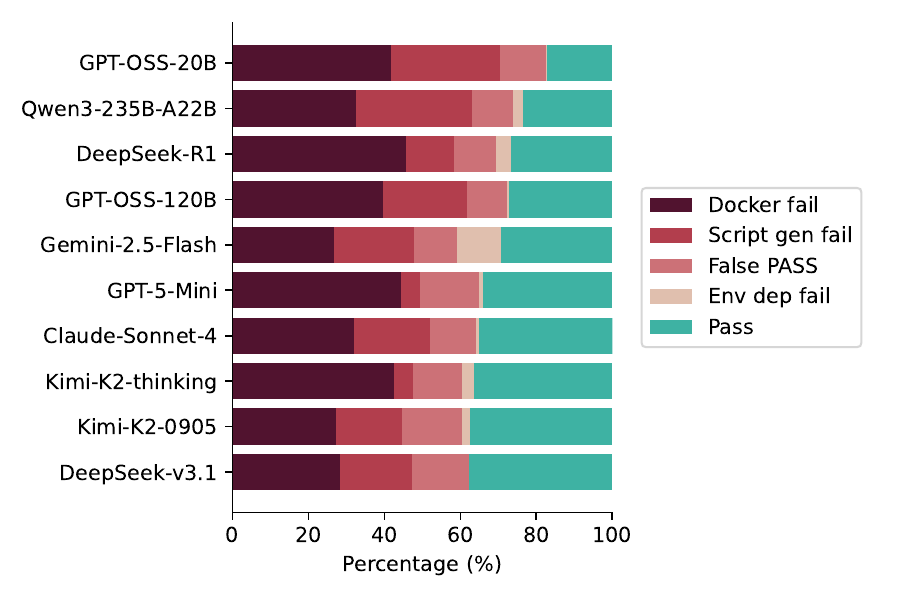}
    \caption{Breakdown of failure modes across models (framework: SWE-Builder). Each bar sums to 100 \%. From left to right: (i) Docker-build failures, (ii) test-script generation failures, (iii) failure to reproduce the reported issue (i.e., the script incorrectly passes on the un-patched code), and (iv) test scripts that cannot run due to missing dependencies in the Docker environment. The first two types of failed agents will not submit answers, while the latter two types of failures occur after submission.}
    \label{fig:fail_reason}
    \vspace{-0.2in}
\end{figure}

To understand what fundamentally limits current systems, we investigate \textit{whether environment construction or test script generation is the primary bottleneck in configuration tasks (RQ4)}. Figure~\ref{fig:fail_reason} reveals a clear hierarchy of failure modes. Docker build errors dominate (avg. 36.09\%), indicating that environment construction is the most failure-prone step. Script-related issues—missing/non-executable scripts (18.12\%) and silent false passes (12.70\%)---form the second major cluster. Once a Docker image is successfully built, downstream evaluation failures are rare (2.63\%), underscoring the robustness of SWE-Builder's self-checking mechanism.

Model-level analysis reinforces this pattern: stronger models reduce script failures but show little improvement on Docker errors, suggesting LLMs lack systematic dependency reasoning. "Thinking-enhanced" models (e.g., \texttt{DeepSeek-R1}, \texttt{GPT-5-Mini}, \texttt{Kimi-K2-thinking}) exemplify this—they produce better scripts but suffer more Docker failures, indicating chain-of-thought reasoning aids code synthesis but not system-level dependency resolution.

\begin{table}[htbp]
\centering
\resizebox{0.5\textwidth}{!}{
\begin{tabular}{lcccc}
\toprule
\textbf{Framework} & \textbf{Metric} & \textbf{Easy (\%)} & \textbf{Hard (\%)} & \textbf{Hard/Easy} \\
\midrule
\multirow{2}{*}{SWE-Builder}
& Commit & 59.35 & 42.55 & 0.72 \\
& \cellcolor{gray!10}Resolved & \cellcolor{gray!10}39.10 & \cellcolor{gray!10}28.40 & \cellcolor{gray!10}0.73 \\
\cmidrule(lr){1-5}
\multirow{2}{*}{RepoLaunch}
& Commit & 47.63 & 16.00 & 0.34 \\
& \cellcolor{gray!10}Resolved & \cellcolor{gray!10}21.71 & \cellcolor{gray!10}5.31 & \cellcolor{gray!10}0.24 \\
\bottomrule
\end{tabular}
}
\caption{Performance comparison on ``Easy'' and ``Hard'' subsets. }
\label{tab:easy_vs_hard}
\vspace{-0.2in}
\end{table}

We further examine whether these bottlenecks persist across frameworks using the "Easy" and "Hard" dataset partitions (Section~\ref{sec:data_collect}). As shown in Table~\ref{tab:easy_vs_hard}, both frameworks achieve higher success on "Easy" cases, though still far from perfect—confirming that agents often struggle with script generation even when dependencies are simple. Under "Hard" conditions, RepoLaunch shows a sharper performance drop than SWE-Builder, highlighting differential resilience to dependency complexity.

These findings collectively demonstrate that \textbf{environment construction, particularly dependency resolution, remains the least stable component—making it the critical target for future agent-based configuration systems.}

\section{Conclusion}

In this work we introduce Multi-Docker-Eval, the first multilingual benchmark that jointly assesses LLM agents' ability to generate runnable Docker environments and valid test scripts for real-world repositories. Through experiments across diverse models and frameworks, we revealed critical insights into the efficiency, robustness, and scalability of current approaches, highlighted that framework architecture and ecosystem standardization play an important role in determining success, underscoring the need for memory-augmented, feedback-driven, and language-aware designs. We hope Multi-Docker-Eval serves as a foundation for advancing fully automated, resource-efficient pipelines in the era of data-driven software engineering.


\bibliography{reference}

\newpage
\onecolumn
\begin{appendix}

\section{Schematic Illustrations of SWE-Builder and RepoLaunch}
\label{app:framework}

\begin{figure}[htbp]
    \centering
    \begin{subfigure}[b]{0.98\linewidth}
      \centering
      \includegraphics[width=\linewidth]{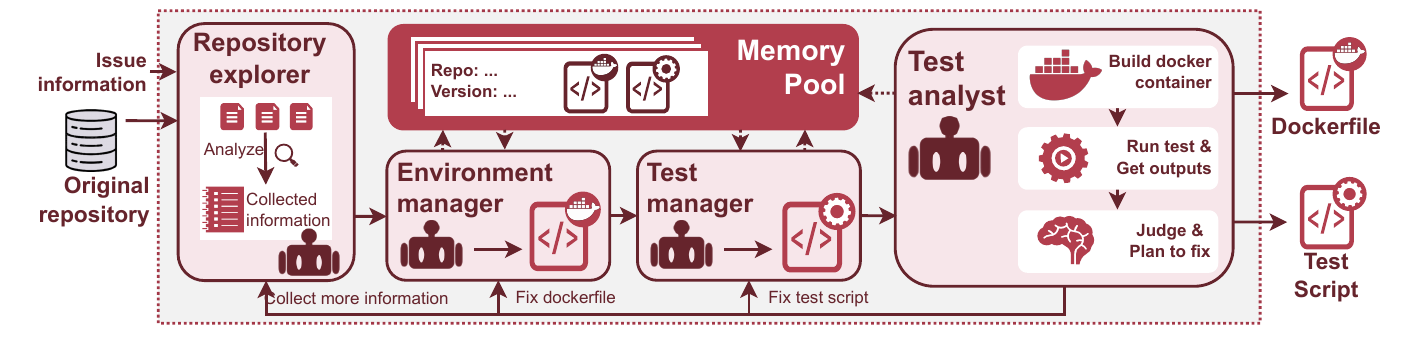}
      \caption{Framework overview of SWE-Builder}
      \label{fig:swe_builder}
    \end{subfigure}\hfill
    \begin{subfigure}[b]{0.98\linewidth}
      \centering
      \includegraphics[width=\linewidth]{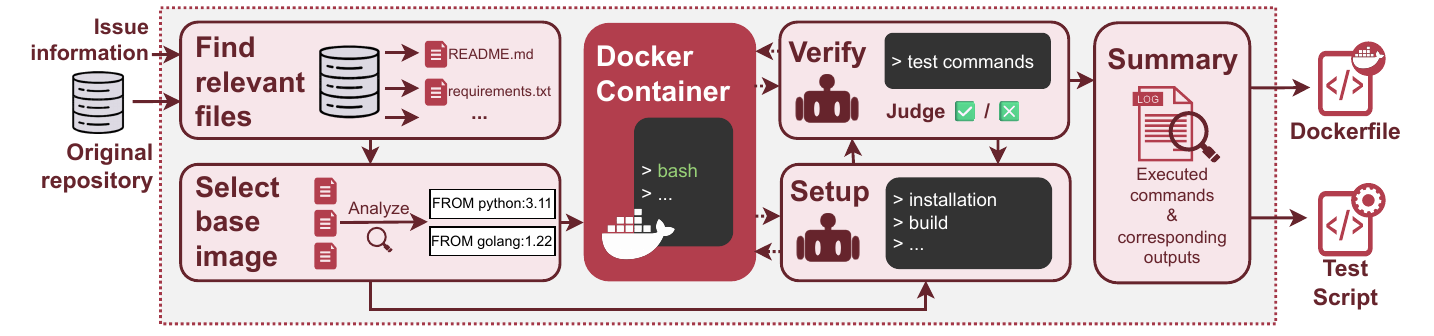}
      \caption{Framework overview of RepoLaunch}
      \label{fig:repolaunch}
    \end{subfigure}
    
    \caption{Overview of the data composition of Multi-Docker-Eval.The framework diagrams of SWE-Builder and RepoLaunch are shown below. Both systems take a source code repository and issue information as input, and produce a Dockerfile along with a test script as output. These outputs collectively provide an executable environment for the repository and enable corresponding tests for the given issue.}
    \label{fig:frameworks}
    \vspace{-0.15in}
\end{figure}

\begin{itemize}
    \item SWE-Builder~\citep{swe_factory} (Figure~\ref{fig:swe_builder}): A multi-agent, iterative framework that partitions the environment setup into four specialized roles: \textit{Repository Explorer}, \textit{Environment Manager}, \textit{Test Manager}, and \textit{Test Analyst}. These agents collaborate to gather dependencies, generate Dockerfiles and test scripts, analyze failures, and iteratively refine configurations. Notably, if the final \textit{Test Analyst} phase yields unsatisfactory results, the system reactivates the other three agents for error correction, forming a closed-loop iterative workflow. To further improve efficiency, SWE-Builder incorporates an \textit{Evaluation Environment Memory Pool} that reuses previously validated configurations from similar repository versions. This multi-agent, memory-augmented design enables robust performance across diverse languages and repositories, making SWE-Builder our primary experimental framework due to its high reliability and scalability.
    \item RepoLaunch~\citep{repolaunch} (Figure~\ref{fig:repolaunch}): A single-agent, sequential workflow. It scans the repository structure, identifies configuration documents, selects a base image, and then launches a Docker container to perform iterative bash commands for dependency installation and test execution. Although local cycles exist between the \textit{Setup} and \textit{Verify} phases, the overall process advances sequentially from one stage to the next. In our extended version, we automated two originally manual steps: enabling the agent to review execution history to extract minimal installation and test commands, and performing automated golden-patch validation using language-specific log parsers. RepoLaunch operates in a vast action space of bash commands, requiring dynamic interaction with Docker containers, which introduces higher uncertainty.
\end{itemize}

\section{Benchmark Details}

\subsection{Repositories in Multi-Docker-Eval}
\label{app:repo}

\begin{figure*}[htbp]
\centering
\begin{minipage}[t]{0.48\textwidth}
\centering
\textbf{Python}
\begin{tasks}[style=itemize](1)
    \task pallets-eco/flask-wtf
    \task rigetti/pyquil
    \task marcelotduarte/cx-Freeze
    \task getlogbook/logbook
    \task pytest-dev/pytest-django
\end{tasks}

\textbf{C}
\begin{tasks}[style=itemize](1)
    \task HandmadeMath/HandmadeMath
    \task libssh2/libssh2
    \task nginx/njs
    \task profanity-im/profanity
\end{tasks}

\textbf{C++}
\begin{tasks}[style=itemize](1)
    \task cpputest/cpputest
    \task GothenburgBitFactory/timewarrior
    \task nfrechette/acl
    \task LiteLDev/LeviLamina
\end{tasks}
\end{minipage}
\hfill
\begin{minipage}[t]{0.48\textwidth}
\centering
\textbf{Go}
\begin{tasks}[style=itemize](1)
    \task uber-go/atomic
    \task warrensbox/terraform-switcher
    \task polarsignals/frostdb
    \task stephenafamo/bob
    \task Altinity/clickhouse-backup
\end{tasks}

\textbf{Java}
\begin{tasks}[style=itemize](1)
    \task OpenAEV-Platform/openaev
    \task java-diff-utils/java-diff-utils
    \task kagkarlsson/db-scheduler
    \task dadoonet/fscrawler
\end{tasks}

\textbf{JavaScript}
\begin{tasks}[style=itemize](1)
    \task pinojs/pino-pretty
    \task prettier/plugin-ruby
    \task vercel/nft
    \task opencomponents/oc
    \task vimeo/player.js
\end{tasks}
\end{minipage}
\caption{The repositories evolved in Multi-Docker-Eval.}
\label{fig:repo}
\end{figure*}

\subsection{Difficulty Verification}
\label{app:diffculty_verf}

\begin{table*}[htbp]
    \centering
    \begin{tabular}{ c | c | m{0.65\textwidth} }
        \toprule
       \centering\textbf{Language}  & \textbf{Base image} & \textbf{Key configuration commands} \\ \midrule
        Python & \makecell{python:3.10-\\slim} & \begin{lstlisting}[ basicstyle=\ttfamily\footnotesize, columns=flexible, aboveskip=1pt, belowskip=-9pt]
RUN if [ -f requirements.txt ]; then pip install -r requirements.txt; fi
RUN if [ -f pyproject.toml ]; then pip install poetry && poetry install || pip install -e .; fi
RUN if [ -f setup.py ]; then pip install -e . || true; fi
RUN if [ -f Pipfile ]; then pip install pipenv && pipenv install --system || true; fi
\end{lstlisting}\\ \hline
    C & gcc:11 & \begin{lstlisting}[ basicstyle=\ttfamily\footnotesize, columns=flexible, aboveskip=1pt, belowskip=-9pt]
RUN if [ -f Makefile ]; then make -j2 || true; else gcc -Wall -O2 $(ls *.c 2>/dev/null) -o main 2>/dev/null || true; fi
\end{lstlisting}
\\ \hline
C++ & gcc:11 & \begin{lstlisting}[ basicstyle=\ttfamily\footnotesize, columns=flexible, aboveskip=1pt, belowskip=-9pt]
RUN if [ -f Makefile ]; then make -j2 || true; else g++ -Wall -O2 $(ls *.cpp 2>/dev/null) -o main 2>/dev/null || true; fi
\end{lstlisting}
\\ \hline
Go & golang:1.20 & \begin{lstlisting}[ basicstyle=\ttfamily\footnotesize, columns=flexible, aboveskip=1pt, belowskip=-9pt]
RUN go install gotest.tools/gotestsum@latest || true
RUN if [ -f go.mod ]; then go mod tidy; fi
RUN go build -v ./... || true
\end{lstlisting}
\\ \hline
JAVA & \makecell{maven:3.9.9-\\eclipse-\\temurin-17} & \begin{lstlisting}[ basicstyle=\ttfamily\footnotesize, columns=flexible, aboveskip=1pt, belowskip=-9pt]
RUN if [ -f pom.xml ]; then mvn -DskipTests package -q; fi
RUN if [ -f build.gradle ]; then gradle build -x test || true; fi
\end{lstlisting}
\\ \hline
JavaScript & \makecell{node:18-\\bullseye-slim} & \begin{lstlisting}[ basicstyle=\ttfamily\footnotesize, columns=flexible, aboveskip=1pt, belowskip=-9pt]
RUN if [ -f package.json ]; then npm install --silent; fi
\end{lstlisting}
\\ \hline
PHP & php:8.1-cli & \begin{lstlisting}[ basicstyle=\ttfamily\footnotesize, columns=flexible, aboveskip=1pt, belowskip=-9pt]
RUN php -r "copy('https://getcomposer.org/installer', 'composer-setup.php');" \
    && php composer-setup.php --install-dir=/usr/local/bin --filename=composer \
    && rm composer-setup.php
RUN composer global require "phpunit/phpunit:^9" --prefer-dist --no-progress --no-suggest || true
ENV PATH="/root/.composer/vendor/bin:$PATH"
RUN if [ -f composer.json ]; then composer install --no-interaction || true; fi
\end{lstlisting}
\\ \hline
Rust & rust:1.70-slim & \begin{lstlisting}[ basicstyle=\ttfamily\footnotesize, columns=flexible, aboveskip=1pt, belowskip=-9pt]
RUN if [ -f Cargo.toml ]; then cargo build --release || true; fi
\end{lstlisting}
\\ \hline
Ruby & ruby:3.1-slim & \begin{lstlisting}[ basicstyle=\ttfamily\footnotesize, columns=flexible, aboveskip=1pt, belowskip=-9pt]
RUN gem install bundler rspec rake minitest test-unit || true
RUN if [ -f Gemfile ]; then bundle install || true; fi
\end{lstlisting}
\\ \bottomrule
    \end{tabular}
    \caption{The base image and key configuration commands utilized to try to build environment. An instance is defined as ``Easy'' if a test-ready environment can be successfully built through these commands. Otherwise, it is assigned the ``Hard'' label.}
    \label{tab:diff_verif}
\end{table*}

\section{Experimental Details}

\subsection{Experimental Platform Configuration}

All experiments were conducted on a virtual machine equipped with 32 CPU cores (Intel® Xeon® Platinum 8457C) and 128 GB of RAM. The system runs Linux version 5.15.0-130-generic with 1000 GB of virtual disk storage. Each experiment was executed in an isolated containerized environment to ensure reproducibility and prevent resource interference across concurrent runs. All evaluations were performed without GPU acceleration.

\subsection{Experimental Parameters}

All parameters not explicitly mentioned here use their framework default values.

\begin{description}
    \item[SWE-Builder Configuration] 
        \quad Concurrency: Concurrency of each round: 5, 8, and 12.
    
    \item[RepoLaunch Configuration] 
        \quad Concurrency: Fixed at 8 concurrent processes for all experiments. \\
        \quad Max steps for setup agent: 30 steps maximum allowed for environment setup tasks. \\
        \quad Max steps for verify agent: 30 steps maximum allowed for verification tasks.
    
    \item[Evaluation Configuration]
        \quad Max workers: 16 worker processes for parallel evaluation. \\
        \quad Docker build timeout: 1800 seconds (30 minutes) maximum allowed for Docker image building. \\
        \quad Test timeout: 2700 seconds (45 minutes) maximum allowed for test execution. \\
        \quad Test runs (for stable F2P): 3 repeated test executions to measure test stability and account for flaky tests.
\end{description}

\section{Detailed Results}

\subsection{Additional Result Figures}
\label{app:more_fig}

\begin{figure}[H]
    \centering
    \begin{minipage}[b]{0.36\linewidth}
        \centering
        \includegraphics[width=\linewidth]{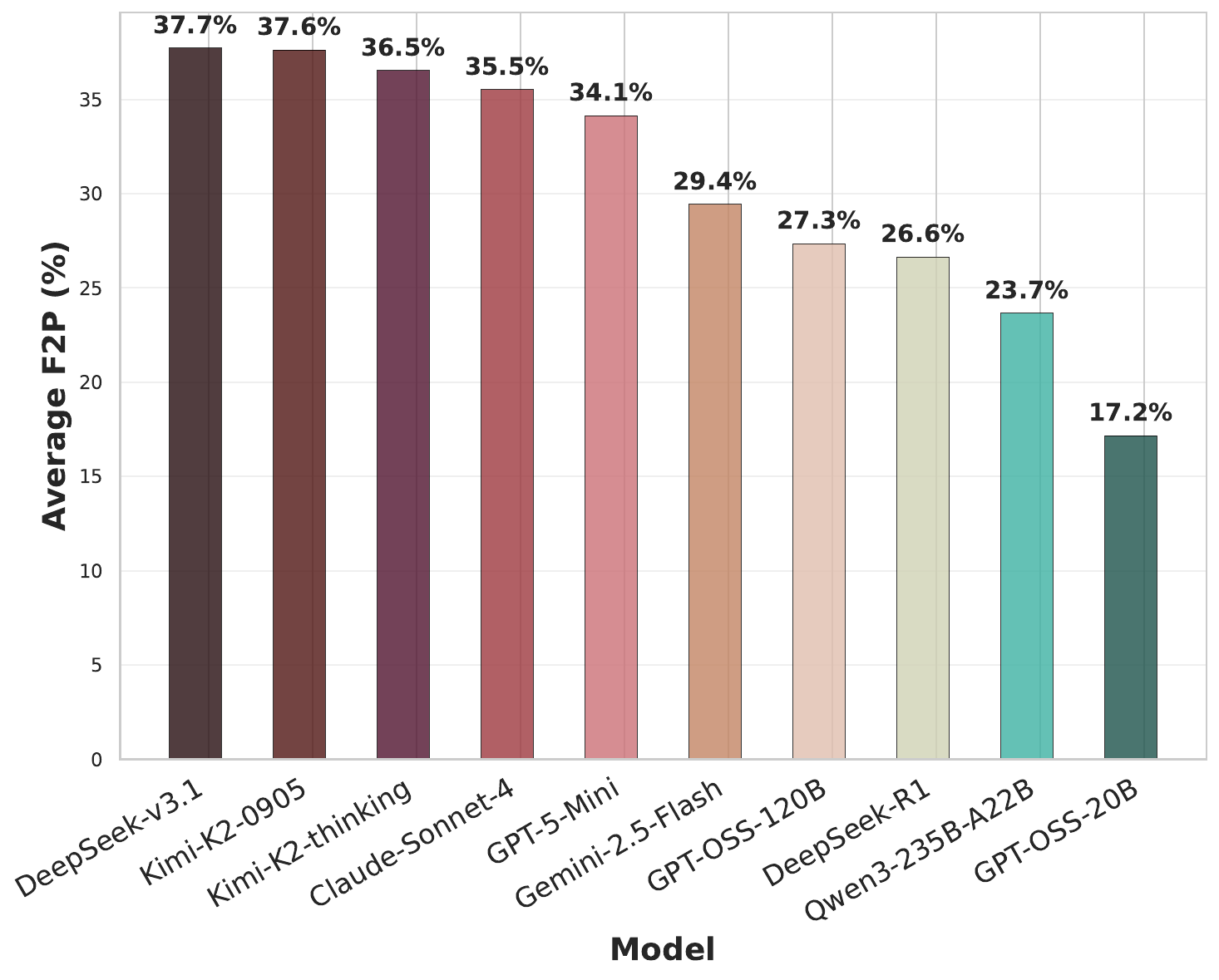}
        \caption{Average F2P of different models (SWE-Builder framework).}
        \label{fig:hist_all}
    \end{minipage}\hfill
    \begin{minipage}[b]{0.6\linewidth}
        \centering
    \includegraphics[width=0.9\linewidth]{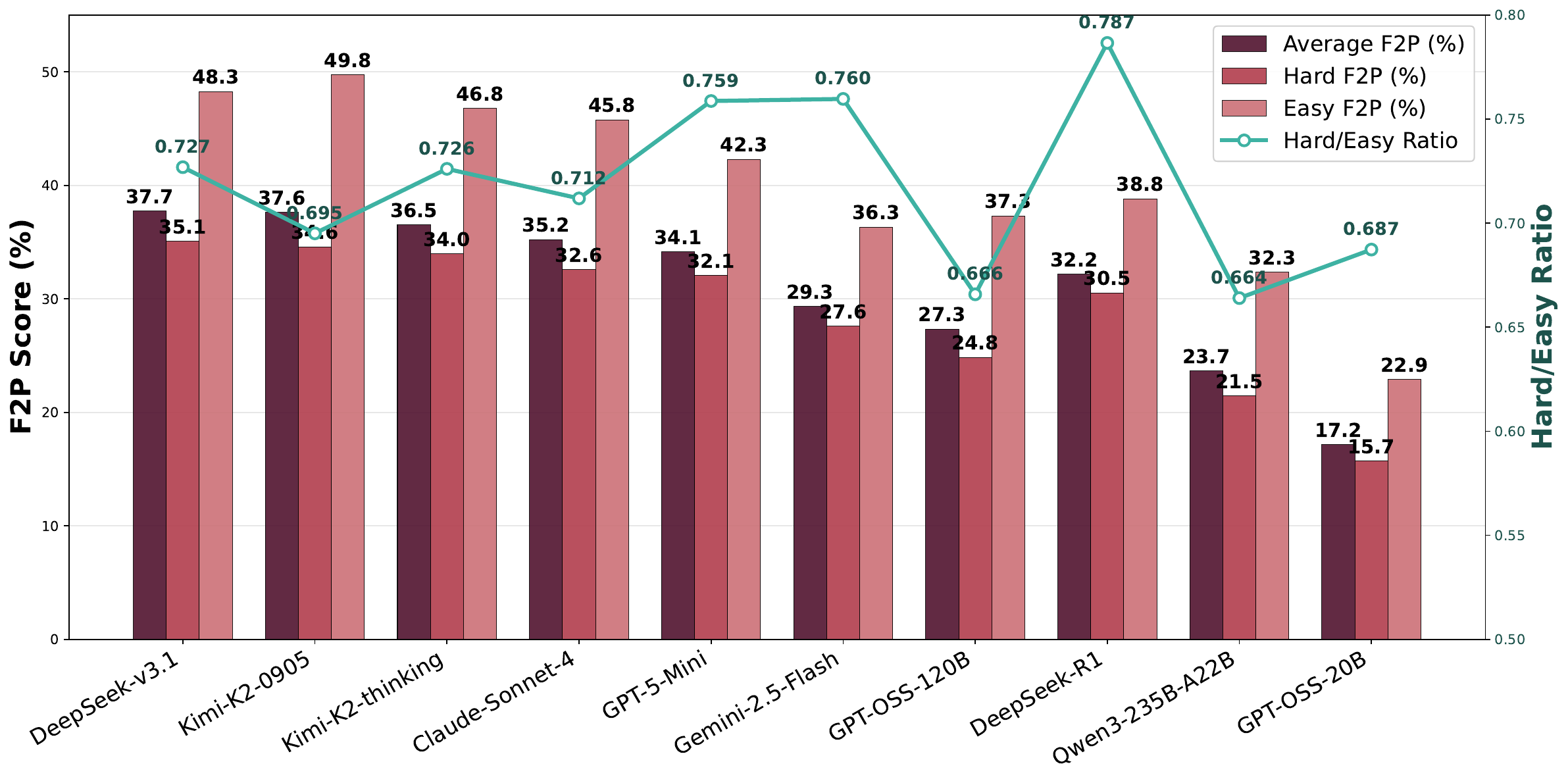}
    \caption{F2P performance across model capabilities: Easy vs. Hard subsets. The line indicates the Hard-to-Easy F2P ratio (Framework: SWE-Builder).}
    \label{fig:easy_vs_hard}
    \end{minipage}
\end{figure}

\subsection{Detailed Results for RepoLaunch Framework}
\label{app:repolaunch}

\begin{table}[H]
  \centering
  \footnotesize
  \setlength{\tabcolsep}{6pt}
  \begin{tabular}{@{}lcccccccc@{}}
    \toprule
    \multicolumn{1}{c}{\multirow{2}{*}{\textbf{Model}}} & \multicolumn{2}{c}{\textbf{Outcome Metrics}} & \multicolumn{6}{c}{\textbf{Process Metrics}} \\
    \cmidrule(lr){2-3} \cmidrule(lr){4-9}
     & \makecell{F2P\\(\%)}  & \makecell{Commit \\ (\%)} & \makecell{Avg \\input\\tokens (K)} & \makecell{Avg \\output\\tokens (K)} & \makecell{Wall \\time\\(Ks)} & \makecell{CPU \\seconds \\ (Ks)} & \makecell{Max\\ RSS\\(GB)} & \makecell{Avg \\docker\\size (GB)} \\
    \addlinespace
    \hline
    \rowcolor{gray!30}\multicolumn{9}{c}{\textit{Open-source Models}} \\
    \texttt{DeepSeek-v3.1} & \textbf{\underline{11.68}} & 35.63 & 197.20 & 1.57 & 22.39 & 1.38 & 5.91 & 1.55 \\
    \rowcolor{gray!10}\texttt{DeepSeek-R1} & 9.88 & 21.56 & 174.35 & 16.02 & 46.33 & 2.04 & 3.78 & 1.39 \\
    \texttt{GPT-OSS-20B} & 2.99 & 6.29 & 202.73 & 1.53 & 22.35 & 1.23 & 6.01 & 1.19 \\
    \rowcolor{gray!10}\texttt{GPT-OSS-120B} & 6.59 & 18.26 & 1690.38 & 5.42 & 26.87 & 1.66 & 6.18 & 1.36 \\
    \texttt{Kimi-K2-0905} & 10.18 & 34.13 & 137.27 & 1.97 & 21.20 & 1.02 & 3.71 & 1.54 \\
    \rowcolor{gray!10}\texttt{Kimi-K2-thinking} & 8.68 & 18.26 & 434.50 & 13.63 & 24.14 & 1.61 & 3.66 & 1.44 \\
    \texttt{Qwen3-235B-A22B} & 3.59 & 6.59 & 111.33 & 10.51 & 26.22 & 0.90 & 7.54 & 1.42 \\
    \addlinespace
    \hline
    \rowcolor{gray!30}\multicolumn{9}{c}{\textit{Closed-source Models}} \\
    \texttt{Claude-Sonnet-4} & 9.88 & 31.74 & 860.02 & 30.62 & 52.21 & 49.56 & 7.81 & 1.49 \\
    \rowcolor{gray!10}\texttt{GPT-5-Mini} & \textbf{\underline{10.18}} & 23.05 & 694.93 & 7.36 & 39.02 & 6.61 & 4.02 & 1.43 \\
    \texttt{Gemini-2.5-Flash} & 9.88 & 22.75 & 859.84 & 12.45 & 27.66 & 1.40 & 6.98 & 1.437 \\
    \bottomrule
  \end{tabular}
  \caption{Overall Multi-Docker-Eval performance of different models on RepoLaunch.}
  \label{tab:repolaunch_result}
  \vspace{-0.15in}
\end{table}

\subsection{F2P Rates Across Programming Languages}
\label{app:language}

\begin{table}[H]
\centering
\resizebox{\textwidth}{!}{
\begin{tabular}{c|*{9}{c}}
\toprule
\multirow{2}{*}{\textbf{Model}} & \multicolumn{9}{c}{\textbf{Language}} \\
\cline{2-10}
 & Python & JavaScript & Java & C++ & C & Go & Ruby & Rust & PHP \\ \hline
\texttt{DeepSeek-v3.1} & 47.86 & 45.83 & \textbf{\underline{14.29}} & \textbf{\underline{18.89}} & 33.33 & 57.50 & \textbf{\underline{51.67}} & 20.83 & \textbf{\underline{42.22}} \\
\rowcolor{gray!10}\texttt{DeepSeek-R1} & 32.48 & 40.83 & 9.52 & 10.00 & 22.50 & 46.67 & 35.83 & 10.00 & 25.56 \\
\texttt{GPT-OSS-20B} & 15.38 & 17.50 & 8.57 & 2.22 & 15.00 & 46.67 & 25.00 & 9.17 & 7.78 \\
\rowcolor{gray!10}\texttt{GPT-OSS-120B} & 35.90 & 36.67 & 10.48 & 10.00 & 23.33 & 53.33 & 32.50 & 14.17 & 22.22  \\
\texttt{Kimi-K2-0905} & 47.01 & \textbf{\underline{49.17}} & 12.38 & 12.22 & 34.17 & 61.67 & \textbf{\underline{51.67}} & \textbf{\underline{22.50}} & 36.67 \\
\rowcolor{gray!10}\texttt{Kimi-K2-thinking} & \textbf{\underline{49.57}} & 46.67 & 11.43 & 11.11 & \textbf{\underline{37.50}} & 63.33 & 48.33 & 17.50 & 32.22 \\
\texttt{Qwen3-235B-A22B} & 21.37 & 37.50 & 9.52 & 5.56 & 17.50 & 51.67 & 30.00 & 15.83 & 15.56 \\
\rowcolor{gray!10}\texttt{Claude-Sonnet-4} & 41.88  & 48.33 & 8.57 & \textbf{\underline{18.89}} & 32.50 & \textbf{\underline{66.67}} & 43.33 & 18.33 & 30.00 \\
\texttt{GPT-5-Mini} & 43.59 & \textbf{\underline{49.17}} & \textbf{\underline{14.29}} & 15.56 & 29.17 & 48.33 & 49.17 & 16.67 & 34.44 \\
\rowcolor{gray!10}\texttt{Gemini-2.5-Flash} & 35.04 & 42.50 & 8.57 & 10.00 & 26.67 & 49.17 & 41.67 & 14.17 & 28.89 \\ \bottomrule
\end{tabular}
}
\caption{F2P Rates Across Programming Languages by Model on SWE-Builder (\%)}
\label{tab:language_model}
\end{table}

\end{appendix}

\end{document}